\documentstyle[11pt]{article}
\newcommand{\beq}{\begin{equation}}
\newcommand{\eeq}{\end{equation}}
\newcommand{\bdm}{\begin{displaymath}}
\newcommand{\edm}{\end{displaymath}}
\newcommand{\beqr}{\begin{eqnarray}}
\newcommand{\eeqr}{\end{eqnarray}}

\begin{document}

\title{A perturbative approach to the quantum elliptic Calogero-Sutherland model}

\author{J. Fern\'andez N\'u\~{n}ez, W. Garc\'{\i}a Fuertes,  A.M. 
Perelomov\footnote{On leave of absence from the Institute for Theoretical and Experimental Physics, 117259, Moscow, Russia. Current E-mail address: perelomo@dftuz.unizar.es}\\     {\small\em Departamento de F\'{\i}sica, Facultad de Ciencias, Universidad de Oviedo, E-33007 Oviedo, Spain}} 

\date{}

\maketitle

\begin{abstract}

We solve perturbatively the quantum elliptic Calogero-Sutherland model in the regime in which  the quotient between the real and imaginary semiperiods of the Weierstrass ${\cal P}$ function is small.

\end{abstract}

\section*{}

The class of quantum and classical integrable systems known as Calogero-Sutherland models were 
first introduced by these authors in the seventies, \cite{ca71,su72}, and have since then attracted considerable interest, both for their intrinsic mathematical beauty and depth and for the numerous applications found, which range from condensed matter to supersymmetric Yang-Mills theory and strings/M-theory, see for instance \cite{appl}. These models, whose integrability stems from the fact that their Hamiltonians coincide with the Laplace-Beltrami operators on some symmetric spaces, can be formulated for an arbitrary number of particles. There are five possible interaction potentials: (a) $V(q)=q^{-2}$; (b) $V(q)=\sinh^{-2}q$; (c) $V(q)=\sin^{-2}q$; (d) $V(q)={\cal P}(q)$, $\cal P$ being the elliptic Weierstrass function; and (e) $V(q)=q^{-2}+\omega^2q^2$. In all cases, the particle coordinates enter in these potentials in combinations which are given by the roots of some simple Lie algebra, see \cite{op83} for details.

The most general among these systems is the elliptic one: all the other potentials arise as 
suitable infinite limits of one of both semiperiods of the ${\cal P}$-function. Nevertheless, to solve the quantum elliptic Calogero-Sutherland model is, even in the most simple cases, a difficult task. The elliptic problem for only one particle and special values of the coupling constant was solved for Lam\'e more than one century ago in the course of his analysis of the stationary distribution of temperatures on an ellipsoid \cite{la}, and later in greater generality by Hermite \cite{he80}. Apart from this, one of the most successful results obtained so far is the exact solution for the case of three particles given in \cite{di93, in95}. However, the final formula for the eigenvalues is a very complicated expression involving transcendental functions, and it is therefore quite hard to grasp its content. In this letter, we will show that in some cases it is
possible to take advantage of the solutions of the trigonometric Calogero-Sutherland model
developed in \cite{pe98a,prz98,pe99} to give approximate solutions to the elliptic problem. These solutions are expressed by simple rational functions and the procedure for finding them is fairly elementary.

We begin by recalling some basic facts taken from \cite{pe98a,prz98,pe99}. The trigonometric 
quantum Calogero-Sutherland model of $A_n$-type describes the mutual interaction of $N=n+1$
particles moving on the circle. The coordinates of these particles are $q_j$, $j=1,\ldots,N$, and the Schr\"{o}dinger equation reads
\beqr
H^{\rm trig}\Psi^\kappa&=&E^{\rm trig}(\kappa)\Psi^\kappa\nonumber\\
H^{\rm trig}=-\frac{1}{2}\Delta&+&\kappa (\kappa-1)\sum_{j<k}^N\sin^{-2}(q_j-q_k),\ \ \ \Delta=
\sum_{j=1}^N\frac{\partial^2}{\partial q_j^2}.\label{1}
\eeqr

The quantum eigenstates depend on a $n$-tuple of quantum numbers ${\bf m}=(m_1,m_2,\ldots,m_n)$
\beqr
H^{\rm trig}\Psi^\kappa_{\bf m}&=&E_{\bf m}^{\rm trig}(\kappa) \Psi_{\bf m}^\kappa \nonumber\\
E_{\bf m}^{\rm trig}(\kappa)&=&2 (\lambda+\kappa\rho,\lambda+\kappa\rho),\label{105}
\eeqr
where $\lambda$ is the highest weight of the representation of $A_n$ labelled by ${\bf m}$, i. e.  $\lambda=\sum_{i=1}^n m_i \lambda_i$ with $\lambda_i$ the fundamental weights of $A_n$, and $\rho$ is the standard Weyl vector, $\rho=\frac{1}{2}\sum_{\alpha\in{\cal R}^+}\alpha$,
with the sum extended over all the positive roots of $A_n$. The center-of-mass-frame eigenfunctions are of the form
\beq
\Psi_{\bf m}^\kappa(q_i)=\Big\{\prod_{j<k}^N\sin(q_j-q_k)\Big\}^\kappa P_{\bf m}^\kappa(z_i),
\eeq
where the $z_i, i=1,2,\ldots,n$, variables are the elementary symmetric functions of 
$x_j=e^{2 i q_j}$, 
\beq
z_p=\sum_{j_1<j_2<\ldots<j_p}^N x_{j_1}x_{j_2}\ldots x_{j_p},
\eeq
and $P_{\bf m}^\kappa$ are the generalized Gegenbauer polynomials related to $A_n$. Some 
properties of these polynomials, as well as explicit examples, can be found in
\cite{pe98a,prz98,pe99,flp01,fp02}. We mention, in particular, that each product of the form $z_iP_{\bf m}$ can be decomposed as a linear combination of Gegenbauer polynomials which mimics the structure of the Clebsch-Gordan series for the irreducible representations of $SU(n)$. Here we only quote two of these recurrence relations which are specially relevant for what follows:
\begin{eqnarray}
z_1 P_{\bf m}^\kappa&=&\sum_{j=1}^{N} c_{j,{\bf m}}^\kappa P^\kappa_{{\bf m}+
{\bf \mu}_j}\nonumber\\
z_n P_{\bf m}^\kappa&=&\sum_{j=1}^{N} \tilde{c}_{j,{\bf m}}^\kappa P_{{\bf m}-{\bf \mu}_j}^\kappa ;
\label{eq:rec}
\end{eqnarray}
in these formulas ${\bf \mu}_j$ is the $n$-tuple whose $i$-th element is $\delta_{i,j}-
\delta_{i,j-1}$ and $c_{j,{\bf m}}^\kappa,\ \tilde{c}_{j,{\bf m}}^\kappa$ are some coefficients
which can be obtained by known algorithms, see \cite{la89, st89, ma95}. Note that as
$z_n=z_1^\dagger$ both recurrence relations are simply related.

The elliptic model related to $A_n$ has the same structure. The Schr\"{o}dinger equation is
$H^{\rm ell}\Phi^\kappa=E^{\rm ell}(\kappa)\Phi^\kappa$, the Hamiltonian being
\beq
H^{\rm ell}=-\frac{1}{2}\Delta+\kappa (\kappa-1)\sum_{j<k}^N{\cal P}(q_j-q_k; \omega_1,\omega_2), 
\eeq
where ${\cal P}(z;\omega_1,\omega_2)$ is the Weierstrass elliptic function with semiperiods chosen 
to be  $\omega_1=\frac{\pi}{2}$ and $\omega_2$ an imaginary number. This ensures that the ${\cal P}$ function will take real values on the real axis. For $|\omega_2|\ll\omega_1$, the Weierstrass function can be expanded in the parameter $g=e^{-4|\omega_2|}$:
\beq
{\cal P}(z;\frac{\pi}{2},\frac{\ln g}{4i})=\sin^{-2}z-\frac{1}{3} +8\sum_{k=1}^\infty \frac{k
g^k}{1-g^k}  (1-\cos 2 k z);
\eeq
that is, $\cal P$ is represented by the explicit power series
\beq
{\cal P}(z;\frac{\pi}{2},\frac{\ln g}{4i})=\sin^{-2}z-\frac{1}{3} +\sum_{p=1}^\infty {g^p}V_p(z),
\eeq
with
\beq
V_p(z)=8\sum_{h\in D_p}h(1-\cos 2 h z),
\eeq
$D_p$ being the set of natural divisors of $p$, i.e., $D_p=\{h\in{\bf N}\,|\,p/h \in{\bf N}\}$.
Therefore, 
\beq
H^{\rm ell}=H^{\rm trig}-\frac{1}{6}\kappa(\kappa-1)N(N-1)+\kappa(\kappa-1)\sum_{p=1}^\infty g^p 
\Big(\sum_{j<k}^N V_p(q_j-q_k)\Big)\label{eq:pert}
\eeq
and a perturbative treatment of the elliptic problem becomes feasible. The first order term in that 
expansion is
\begin{eqnarray}
\kappa(\kappa-1) g \sum_{j<k}^N V_1(q_j-q_k)=8g\kappa(\kappa-1) \sum_{j<k}^N\big(1-
\cos 2(q_j-q_k)\big)=4g\kappa (\kappa-1)  (N^2-z_1z_n),
\end{eqnarray}
and thus, first order perturbation theory gives
\begin{eqnarray}
E_{\bf m}^{\rm ell}(\kappa)&=&E_{\bf m}^{\rm trig}(\kappa)-\frac{1}{6}\kappa(\kappa-1)N(N-1)+
\delta_1 E_{\bf m}(\kappa)+o(g^2)\label{eq:en}\\
\delta_1 E_{\bf m}(\kappa)&=&4g\kappa(\kappa-1)\frac{\langle \Psi_{\bf m}^\kappa|N^2-z_1z_n|
\Psi_{\bf m}^\kappa\rangle}{\langle \Psi_{\bf m}^\kappa|\Psi_{\bf m}^\kappa\rangle}\label{eq:c1}.
\end{eqnarray}

The recurrence relations for the trigonometric case allow an easy evaluation of the energy 
correction: it follows from (\ref{eq:rec}) that
\beq
z_1 z_n P_{\bf m}^\kappa=a_{\bf m}^\kappa P_{\bf m}^\kappa+\cdots ,
\eeq
where the dots stand for terms proportional to polynomials other than $P_{\bf m}^\kappa$ and
\beq
a_{\bf m}^\kappa=\sum_{j=1}^N  \tilde{c}_{j,{\bf m}}^\kappa  c_{j,{\bf m}-\bf{\mu}_j}^\kappa\,.
\eeq
The orthogonality properties of the system of generalized Gegenbauer polynomials guarantee that 
these terms do not contribute to (\ref{eq:c1}), and we come to the simple result
\beq
\delta_1 E_{\bf m}(\kappa)=4g\kappa(\kappa-1)\left[N^2- a_{\bf m}^\kappa \right].
\eeq

We can use the explicit expression of the coeffients $c_{j,{\bf m}}^\kappa,\ 
\tilde{c}_{j,{\bf m}}^\kappa$ given in \cite{pe99, flp01} to write this correction for the
$A_1,A_2$ and $A_3$ cases, i.e. for two, three and four particles, respectively.\\

\noindent$\bullet$ $A_1$ case: Here ${\bf m}=(m),\ N=2$ and 

\beq
\begin{array}{lcl}
c_{1,m}^\kappa=1&\ \ \ &c_{2,m}^\kappa=c_m(\kappa)\\
\tilde{c}_{1,m}^\kappa=c_m(\kappa)&\ \ \ &\tilde{c}_{2,m}^\kappa=1\label{eq:cm}
\end{array}
\eeq
with
\beq
c_{m}(\kappa)=\frac{m(m-1+2\kappa)}{(m+\kappa)(m-1+\kappa)}\label{eq:a1ce}.
\eeq

Therefore
\beq
\delta_1 E_{m}(\kappa)=8 g \kappa (\kappa-1)\left[ 1+\frac{\kappa (\kappa-1)}
{(m+1+\kappa)(m-1+\kappa)}\right].
\eeq

\noindent$\bullet$ $A_2$ case: Here ${\bf m}=(m,n),\ N=3$ and 

\beq
\begin{array}{lclcl}
c_{1,(m,n)}^\kappa=1&\ \ \ &c_{2,(m,n)}^\kappa=c_m(\kappa)&\ \ \ &c_{3,(m,n)}^\kappa=
a_{m,n}(\kappa)\\
\tilde{c}_{1,(m,n)}^\kappa=a_{n,m}(\kappa)&\ \ \ &\tilde{c}_{2,(m,n)}^\kappa=
c_n(\kappa)&\ \ \ &\tilde{c}_{3,(m,n)}^\kappa=1
\end{array}
\eeq
with
\beq
a_{m,n}(\kappa)=\frac{n(m+n+\kappa)(n-1+2\kappa)(m+n-1+3\kappa)}
{(n+\kappa)(n-1+\kappa)(m+n+2\kappa)(m+n-1+2\kappa)}.
\eeq
Therefore
\begin{eqnarray}
\delta_1 E_{m,n}(\kappa)&=&24 g \kappa (\kappa-1)
+8g\kappa^2(\kappa-1)^2\,\frac{3\kappa^2+3(m+n)\kappa+m^2+n^2+mn-3}
{(m+1+\kappa)(m-1+\kappa)(n+1+\kappa)}\nonumber\\&\times&
\frac{2\kappa^2+(3m+3n+1)\kappa+m^2+n^2+mn-1}{(n-1+\kappa)(m+n+1+2\kappa)(m+n-1+2\kappa)}.
\end{eqnarray}

\noindent$\bullet$ $A_3$ case: Here ${\bf m}=(m,l,n),\ N=4$ and 
\beq
\begin{array}{lclclcl}
c_{1,(m,l,n)}^\kappa=1&\ \ \ &c_{2,(m,l,n)}^\kappa=c_m(\kappa)&\ \ \ &c_{3,(m,l,n)}^\kappa=
a_{m,l}(\kappa)&\ \ \ &c_{4,(m,l,n)}=d_{m,l,n}(\kappa)\\
\tilde{c}_{1,(m,l,n)}^\kappa=d_{n,l,m}(\kappa)&\ \ \ &\tilde{c}_{2,(m,l,n)}^\kappa=
a_{n,l}(\kappa)&\ \ \ &\tilde{c}_{3,(m,l,n)}^\kappa=c_n(\kappa)&\ \ \
&\tilde{c}_{4,(m,l,n)}^\kappa=1
\end{array}
\eeq
with
\beq
d_{m,l,n}(\kappa)=\frac{n(l+n+\kappa)(n-1+2\kappa)(m+l+n+2\kappa)(l+n-1+3\kappa)(m+l+n-1+4\kappa)}
{(n+\kappa)(n-1+\kappa)(l+n+2\kappa)(l+n-1+2\kappa)(m+l+n+3\kappa)(m+l+n-1+3\kappa)}.
\eeq

Therefore 
\begin{eqnarray}
\delta_1 E_{m,l,n}(\kappa)&=&4 g \kappa(\kappa-1)\nonumber\\
&\times&\left[16-\frac{n(l+1)(l+m+1+\kappa)(l+2\kappa)(n-1+2\kappa)(l+m+3\kappa)}
{(l+\kappa)(l+1+\kappa)(n+\kappa)(n-1+\kappa)(l+m+2\kappa)(l+m+1+2\kappa)}\right.\nonumber\\
&-&\frac{(n+1)(l+n+1+\kappa)(n+2\kappa)(l+m+n+1+2\kappa)(l+n+3\kappa)(l+m+n+4\kappa)}
{(n+\kappa)(n+1+\kappa)(l+n+2\kappa)(l+n+1+2\kappa)(l+m+n+3\kappa)(l+m+n+1+3\kappa)}\nonumber\\
&-&\frac{m(l+m+\kappa)(m-1+2\kappa)(l+m+n+2\kappa)(l+m-1+3\kappa)(l+m+n-1+4\kappa)}
{(m+\kappa)(m-1+\kappa)(l+m+2\kappa)(l+m-1+2\kappa)(l+m+n+3\kappa)(l+m+n-1+3\kappa)}\nonumber\\
&-&\left.\frac{l(m+1)(l+n+\kappa)(m+2\kappa)(l-1+2\kappa)(l+n-1+3\kappa)}
{(l+\kappa)(l-1+\kappa)(m+\kappa)(m+1+\kappa)(l+n+2\kappa)(l+n-1+2\kappa)}\right].
\end{eqnarray}

This expression becomes particularly simple when only one quantum number is non-vanishing:
\begin{eqnarray}
\delta_1 E_{m,0,0}(\kappa)&=&24 g
\kappa(\kappa-1)\Big[2+\frac{4\kappa^3+(4m-2)\kappa^2+(m^2-2)\kappa}
{(m-1+\kappa)(1+2\kappa)(m+1+3\kappa)}\Big]\nonumber\\
\delta_1 E_{0,l,0}(\kappa)&=&16 g \kappa(\kappa-1)\Big[3+\frac{3\kappa^3+4l\kappa^2+(l^2-3)\kappa}
{(1+\kappa)(l-1+\kappa)(l+1+3\kappa)}\Big]\nonumber\\
\delta_1 E_{0,0,n}(\kappa)&=&24 g
\kappa(\kappa-1)\Big[2+\frac{4\kappa^3+(4n-2)\kappa^2+(n^2-2)\kappa}
{(n-1+\kappa)(1+2\kappa)(n+1+3\kappa)}\Big].
\end{eqnarray}

The extension of the perturbative approach to higher orders is straightforward.
There is only a new ingredient: due to the contribution of intermediate states, the
norms of the unperturbed eigenfunctions enter explicitly in the corrections to the
energies. Fortunately, these norms are known \cite{ma95, op89}. Apart from this,  the procedure to follow is analogous to that used in first  order and the keypoint is that the recurrence relations will save us from doing all the difficult integrals. We will analyse only the simplest example, that
is, second order perturbation theory for the $A_1$ case. 

The second order contribution to the Hamiltonian $H^{\rm ell}$ (\ref{eq:pert}) 
\begin{eqnarray}
\kappa(\kappa-1) g^2 \sum_{j<k}^N V_2(q_j-q_k)&=&8 \kappa(\kappa-1) g^2\sum_{j<k}^N[3-
\cos 2(q_j-q_k)-2 \cos 4(q_j-q_k)]\nonumber\\
&=&4\kappa (\kappa-1) g^2[3N^2-z_1 z_n-2 (z_1^2-2z_2)(z_n^2-2z_{n-1})]
\end{eqnarray}
for the $A_1$ case  gives, taking $z_0=1$,
\beq
E_{m}^{\rm ell}(\kappa)=E_{m}^{\rm trig}(\kappa)-\frac{1}{6}\kappa(\kappa-1)N(N-1)+\delta_1
 E_{m}(\kappa)+ \delta_2 E_{m}(\kappa)+o(g^3)
\eeq
with
\begin{eqnarray}
\delta_2 E_{m}(\kappa)&=&4g^2\kappa(\kappa-1)\frac{\langle \Psi_{m}^\kappa|4+7z_1^2-2z_1^4|
\Psi_{m}^\kappa\rangle}{\langle \Psi_{m}^\kappa|\Psi_{m}^\kappa\rangle}\nonumber\\
&+&\frac{16g^2\kappa^2(\kappa-1)^2}{ \langle \Psi_{m}^\kappa|\Psi_{m}^\kappa\rangle}
\sum_{n\neq m}\frac{1}{\langle \Psi_{n}^\kappa|\Psi_{n}^\kappa\rangle}\frac{\left|\langle 
\Psi_{n}^\kappa|z_1^2|\Psi_{m}^\kappa\rangle\right|^2}{E_m^{\rm trig}(\kappa)-E_n^{\rm
trig}(\kappa)}\label{eq:pert2}.
\end{eqnarray}

The recurrence relations for the Gegenbauer polynomials related to $A_1$ (see (\ref{eq:rec}) and
(\ref{eq:cm})) also hold for the eigenfunctions $\Psi_{m}^\kappa$ of the trigonometric problem, i. e., $z_1 \Psi^\kappa_m=\Psi^\kappa_{m+1}+c_m\Psi^\kappa_{m-1}$, from which and the Hermitian
character of $z_1$ (for $A_1$) it follows that
\beq
\langle \Psi_{m}^\kappa|\Psi_{m}^\kappa\rangle=c_m\langle
\Psi_{m-1}^\kappa|\Psi_{m-1}^\kappa\rangle,
\eeq
a very useful relation which allows to express the equation  (\ref{eq:pert2})  in terms of the
coefficients $c_m$ only:
\begin{eqnarray}
\delta_2E_{m}(\kappa)
&=&4g^2\kappa(\kappa-1)[4+7(c_m+c_{m+1})-2(c_m+c_{m+1})^2-2c_{m+2}c_{m+1}-2c_{m}c_{m-1}]
\nonumber\\ &+&16g^2\kappa^2(\kappa-1)^2\left\{\frac{c_{m+2}c_{m+1}}{E_m^{\rm
trig}(\kappa)-E_{m+2}^{\rm trig}(\kappa)}+\frac{c_{m}c_{m-1}}{E_m^{\rm
trig}(\kappa)-E_{m-2}^{\rm trig}(\kappa)}\right\}\label{eq:del2}.
\end{eqnarray}

Finally, using (\ref{eq:a1ce}) and the expression $E_m^{\rm trig}=m^2+2\kappa m-\kappa^2$ for the
energy levels, (\ref{eq:del2})  can be evaluated quite  easily, obtaining  the simple result
\begin{eqnarray}
&&\delta_2 E_{m}(\kappa)=8g^2\kappa^2(\kappa-1)^2\left\{3-\frac{\kappa^2-(10m+6)\kappa-5m^2+8}
{[(m+\kappa)^2-4][(m+\kappa)^2-1]}\right\}\nonumber\\
&+&\frac{4 g^2 \kappa^2(\kappa-1)^2}{m+\kappa}
\left\{\frac{m(m-1)(m-1+2\kappa)(m-2+2\kappa)}{(m-1+\kappa)^3(m-2+\kappa)}-
\frac{(m+1)(m+2)(m+2\kappa)(m+1+2\kappa)}{(m+1+\kappa)^3(m+2+\kappa)}\right\}.\nonumber\\
\end{eqnarray}
\\
{\Large\bf Acnowledgments}\\

We are grateful to Prof. M. Lorente for interesting discussions. A.M.P. would like to express his gratitude to the Department of Physics of the University of Oviedo for the hospitality during  his
stay as a Visiting Professor. The work of J.F.N. and W.G.F. has been partially supported by the
University of Oviedo, Vicerrectorado de Investigaci\'on, grant MB-02-514.\\

\end{document}